\documentclass[10pt,twocolumn,letterpaper]{article}

\usepackage{iccv}              

\definecolor{iccvblue}{rgb}{0.21,0.49,0.74}
\usepackage[pagebackref,breaklinks,colorlinks,allcolors=iccvblue]{hyperref}

\usepackage{multirow} 
\usepackage{makecell}
\usepackage{enumitem}
\usepackage{mathrsfs}
\usepackage{float} 
\usepackage{colortbl}
\usepackage{array}
\usepackage{bm}

\def\paperID{760} 
\def\confName{ICCV}
\def\confYear{2025}

\title{Recover Biological Structure from Sparse-View Diffraction Images\\
with Neural Volumetric Prior}

\author{
Renzhi He\textsuperscript{1}, 
Haowen Zhou\negmedspace\textsuperscript{2},
Yubei Chen,\negmedspace\textsuperscript{1}~~
Yi Xue,\negmedspace\textsuperscript{1}\\
\textsuperscript{1} University of California, Davis
\textsuperscript{2} California Institute of Technology\\
{\tt\small \{cubhe, yxxue, ybchen\}@ucdavis.edu, hzhou7@caltech.edu}
}

\begin{document}
\maketitle
\begin{abstract}
Volumetric reconstruction of label-free living cells from non-destructive optical microscopic images reveals cellular metabolism in native environments. However, current optical tomography techniques require hundreds of 2D images to reconstruct a 3D volume, hindering them from intravital imaging of biological samples undergoing rapid dynamics. This poses the challenge of reconstructing the entire volume of semi-transparent biological samples from sparse views due to the restricted viewing angles of microscopes and the limited number of measurements. 
In this work, we develop Neural Volumetric Prior (NVP) for high-fidelity volumetric reconstruction of semi-transparent biological samples from sparse-view microscopic images. NVP integrates explicit and implicit neural representations and incorporates the physical prior of diffractive optics. We validate NVP on both simulated data and experimentally captured microscopic images. Compared to previous methods, NVP significantly reduces the required number of images by nearly 50-fold and processing time by 3-fold while maintaining state-of-the-art performance.
NVP is the first technique to enable volumetric reconstruction of label-free biological samples from sparse-view microscopic images, paving the way for real-time 3D imaging of dynamically changing biological samples. \href{https://xue-lab-cobi.github.io/Sparse-View-FDT/}{Project Page}\textbf{}
\end{abstract}    
\section{Introduction}
\label{sec:intro}

Tomography is widely used to reconstruct 3D biomedical structures from multiple 2D image views. Computed Tomography (CT), for instance, produces images of the human body based on varying absorption of X-ray by different tissues, while optical tomography produces microscale 3D images of label-free cellular structures based on varying diffraction of visible light by intracellular structures. Numerous advanced optical tomography systems and computational algorithms \cite{Park:06, Kim:17, shaffer2012single, marthy2024single, 9260959, tayal2020simultaneous, 9376596, Liu:18,PHAM2021127290, Choi2007-ay, Sung:09, Waller2010-oj, Tian2015-ij, Chowdhury:17, Yeh:19, Dong2020-dq} have been developed for cultured cells or thin tissue slices, where 3D intracellular structures are reconstructed from diffracted laser or LED light as it transmits through the sample with heterogenous refractive index (RI). However, when the sample is too thick for visible light to transmit through, such as intact, bulky specimens or living animals, these imaging techniques cannot reconstruct 3D structures because the transmission side is inaccessible. 
\begin{figure}[t]
  \centering
   \includegraphics[width=\linewidth, trim={30 500 390 0}, clip]{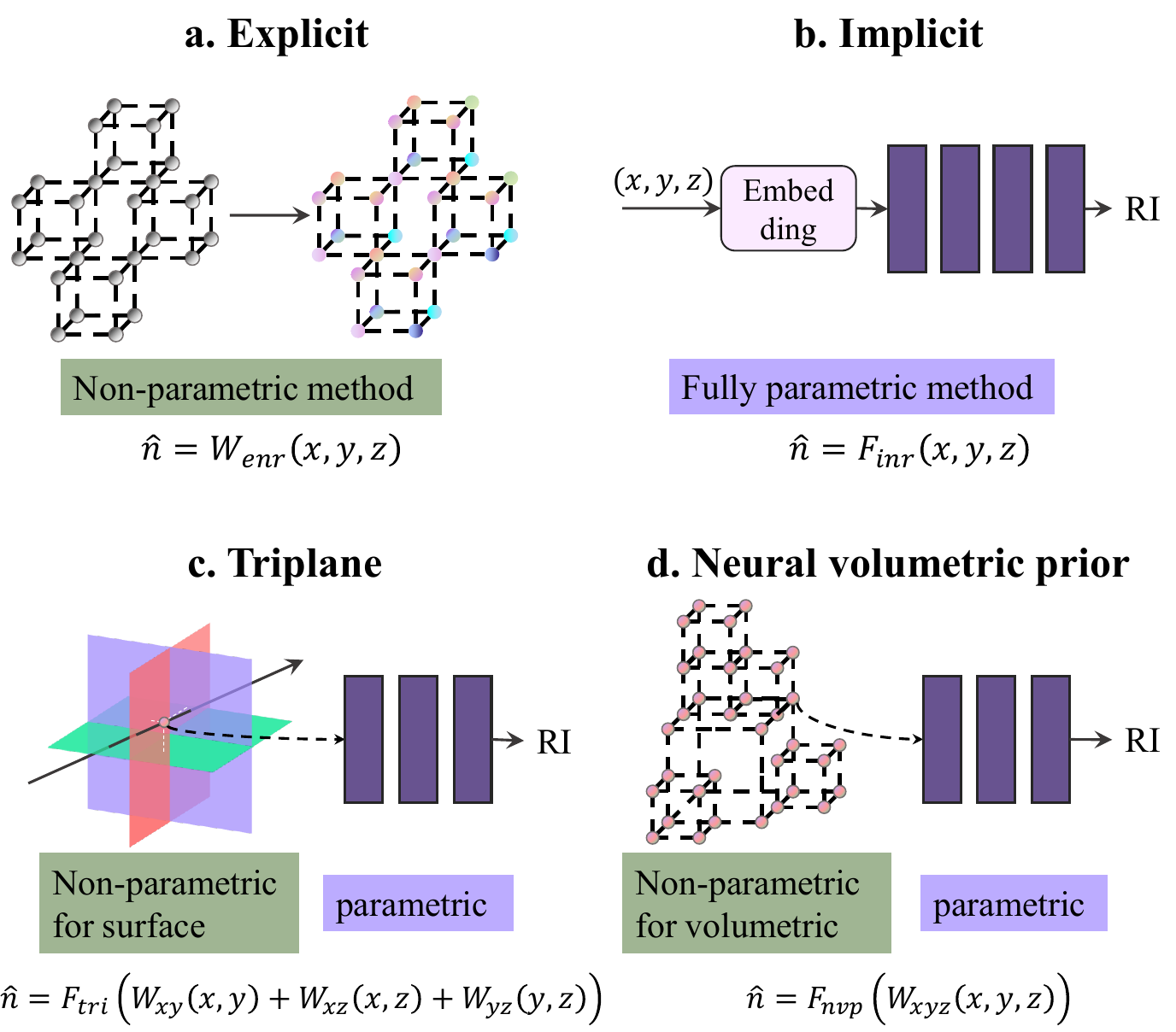}
   \caption{
    \textbf{Illustrations of different neural representation methods.}
    (a) Explicit neural representation: A non-parametric method in which the refractive index (RI) distribution $\hat{n}$ is directly reconstructed by the projection function $W_{\text{enr}}(x, y, z)$, which provide a one-to-one mapping from spatial coordinates to RI values \cite{Xue:22, FDT}. 
    (b) Implicit neural representation: Instead of directly reconstructing $\hat{n}$, this method optimizes the parameters of a  multi-layer perceptron (MLP) $F_{\text{inr}}$, which is then used to predict $\hat{n}$ \cite{decaf}.  
    (c) Triplane: A hybrid method combining non-parametric and parametric components by solving both the triplane features ${W_{xy}, W_{xz}, W_{yz}}$ and the neural model $F_{\text{tri}}$ to reconstruct $\hat{n}$ \cite{kplanes_2023}.  
    (d) NVP: Our proposed hybrid approach reconstructs $\hat{n}$ by integrating an uncompressed volumetric prior ${W_{xyz}}$ into the neural model $F_{\text{nvp}}$.
   }
   \label{fig:NVP}
   \vspace{-7mm}
\end{figure}
To address this challenge, Fluorescence Diffraction Tomography (FDT) \cite{Xue:22, FDT} has been developed as a promising technique for intravital optical tomography. It reconstructs 3D label-free structures from diffracted fluorescence without requiring access to the transmission side, using intrinsic or exogenous fluorescence within the sample as internal light sources. Despite its potential, FDT still requires hundreds of 2D images to reconstruct a 3D volume, requiring that the object remain static during image acquisition (seconds to minutes). This limitation restricts FDT's application in capturing dynamic cellular or tissue structural changes in real time, where high-speed image acquisition is essential \cite{voleti2019real}.

To achieve high-speed volumetric imaging with FDT, it is necessary to reduce the required number of 2D images while preserving the accuracy and resolution of the reconstructed 3D volumes. Compared to 3D rendering of natural scenes, rendering semi-transparent biological samples from diffracted microscopic images is more challenging due to fewer measurements and more unknowns. First, \emph{the range of viewing angles} in microscopy is constrained by the numerical aperture of the system and the fixed position of the camera, unlike natural scenes that can be imaged from full 360-degree perspectives. Second, the spatial distribution of fluorescence sources in real samples further limits \emph{the number of available viewing angles}, resulting in a ``sparse view" setting that complicates volumetric reconstruction. Third, microscopic tomography aims to reconstruct \emph{the entire volume of semi-transparent biological samples}, unlike surface reconstruction of opaque objects in nature scenes, leading to a much larger number of unknown voxels. 

To solve the unique challenges of volumetric rendering with microscopic images, the unknown voxels can be represented by both implicit \cite{nerf} and explicit \cite{Plenoxels} neural fields \cite{Zhou:23,kang2024coordinate, kang2024adaptive,decaf,FDT} (\cref{fig:NVP}a,b) and reconstructed from multiple 2D images. However, previous approaches \cite{FDT} require hundreds of 2D images for the volumetric rendering, limiting their application in reconstructing dynamic structures from sparse views. Even though 3D rendering of the surface of opaque nature scenes from sparse views has been achieved by many methods \cite{sun2022direct, chen2022tensorftensorialradiancefields, instantNPG, Li_2023_CVPR, Wang_2023_ICCV, song2023nerfplayerstreamabledynamicscene}, they cannot directly apply to volumetric reconstruction of semi-transparent biological samples. One potential approach to reduce the requires number of 2D images for biological imaging is to employ the network's structural prior with radial fields \cite{ulyanov2018deep}, such as triplane and hexplane methods \cite{kplanes_2023, Hexplane}, which has been applied for MRI imaging \cite{triplane_MR}. However, because the low-rank decomposition, it can introduce grid artifacts when representing volumetric structures.(\cref{fig:NVP}c). 

In this work, we develop NVP (\cref{fig:NVP}d), a hybrid neural representation model that incorporates explicit neural representations and implicit neural fields as neural volumetric priors to achieve physically accurate volumetric reconstruction of label-free, semi-transparent biological samples from sparse views. The NVP regularize the inverse problem to enable accurate reconstruction of large-scale volumes from only 1.5\% of the 2D images required by previous work \cite{FDT}. The key innovations of our method are:

\begin{enumerate}[label=\roman*)]

    
    
    \item We demonstrate, for the first time, the capability to reconstruct volumetric RI of semi-transparent biological samples from diffracted fluorescence images with limited angles and sparse views, validated through both simulations and real-world experiments. This work opens up a new avenue in diffraction-informed neural volumetric representations.

    \item We demonstrate the effectiveness and efficiency of NVP in optical tomography, reducing the required number of images by nearly 50-fold and processing time by 3-fold compared to previous methods \cite{FDT} in our demonstrated experiments.

    \item We leverage the physical prior of light diffraction to achieve physically accurate rendering and quantitative RI reconstruction of volumetric objects, overcoming the limitations of ray-optics models at microscale imaging and broadening the applicability of neural fields in microscopic imaging.
    
\end{enumerate}

\begin{figure*}[htbp]
  \centering
   \includegraphics[width=0.95\linewidth, trim={0 670 0 0}, clip]{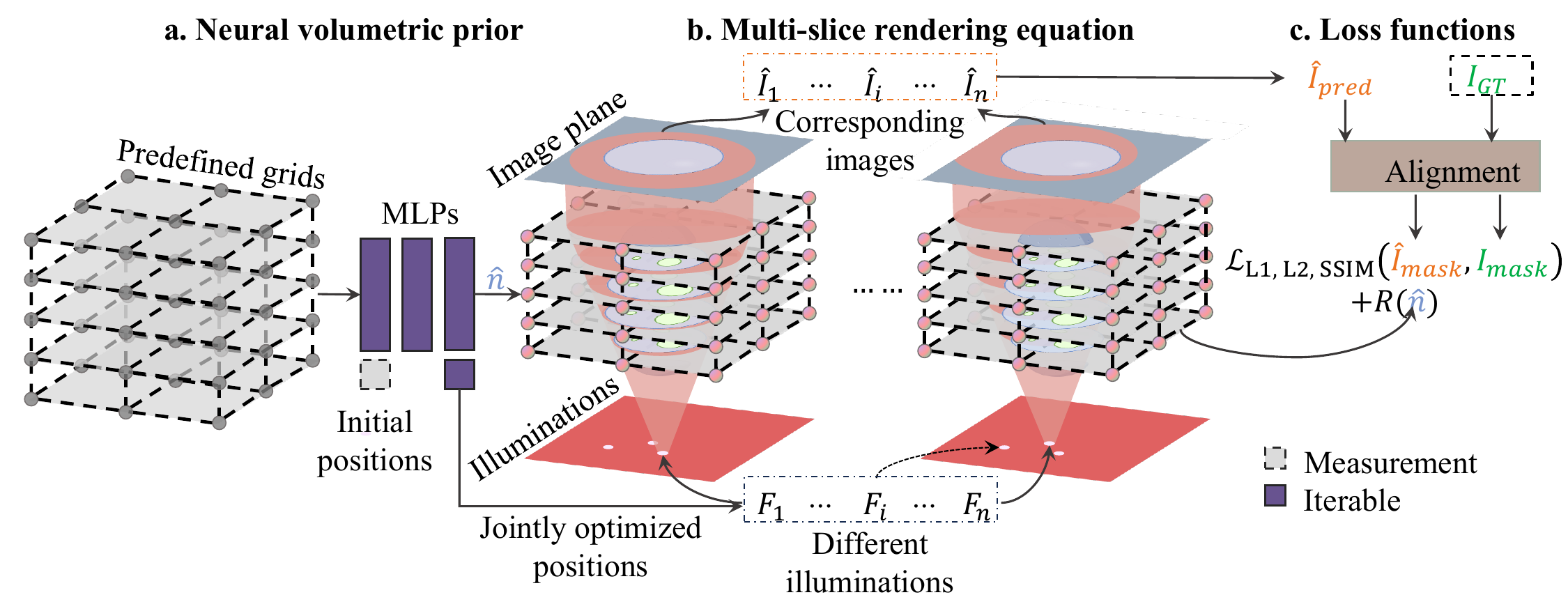}
   \caption{\textbf{Overview of our method for 3D reconstruction.}
    a. Neural volumetric prior (NVP): Predefined 3D grids are reshaped and processed by MLPs to generate the predicted 3D RI volume $\hat{n}$.
    b. Multi-slice rendering equation: The multi-slice model calculates light propagation through the volumetric sample from the fluorescence sources (white spots at the bottom) by accounting for light diffraction at each slice. Each illumination configuration of the fluorescence sources (e.g., $F_1, F_i, F_n$) interacts with the volume to produce a corresponding set of rendered images ($\hat{I}_1, \hat{I}_i, \hat{I}_n$). The illumination configurations are jointly optimized with the RI volume.
    c. Loss functions: The predicted images $\hat{I}_{\text{pred}}$ are aligned with ground truth images $I_{\text{GT}}$ and masked based on light coherence, resulting in $\hat{I}_{\text{mask}}$ and $I_{\text{mask}}$. Loss functions, including $L1$, $L2$, and SSIM, are calculated over the masked regions of $\hat{I}_{\text{pred}}$ and $I_{\text{GT}}$, and a total variation (TV) regularizer $R(\hat{n})$ is applied. 
   }
   \label{fig:overview}
   \vspace{-4mm}
\end{figure*}

\section{Related Work}
\label{sec:formatting}

\subsection{3D Reconstruction from Sparse Viewpoints }

To accurately reconstruct 3D volumes from sparse views, various priors have been explored, such as local geometry \cite{Spurfies,Cai_2024_CVPR, RegNeRF}, multiview consistency \cite{ConsistentNeRF, Zou2023Sparse3D, jain2021puttingnerfdietsemantically, FSGS}, signed distance functions \cite{Long2022SparseNeuS, SparseCraft}, depth regularization \cite{deng2024depthsupervisednerffewerviews,yu2022monosdfexploringmonoculargeometric}, and diffusion-based probabilistic generation \cite{müller2023diffrfrenderingguided3dradiance, liu2023zero1to3zeroshotimage3d}. Even though these approaches produce high-accuracy 3D nature scenes \cite{Yao2023Geometry-Guided, Zhou2021DP-MVS, Zhou2022SparseFusion, MicroDiffusion,sparsect}, 
they model light propagation using ray optics, assuming that light travels in straight lines. 
While this assumption is valid for natural scenes at the macroscale, it becomes inaccurate at the microscale --- where object sizes are comparable to the wavelength of light --- as diffraction effects pose challenges in applying these priors to microscopy. To overcome this challenge, we propose developing a physics-informed rendering equation based on diffraction optics for high-accuracy reconstruction of microscale objects from microscopic images.

\subsection{3D Reconstruction using Diffraction Optics}
Cameras can directly measure the intensity of light but not its phase, whereas heterogeneous RI alters the phase of light. To reconstruct the RI of transparent objects, many computational models based on diffractive optics (i.e., wave optics) have been developed. The most classic model is the Born approximation \cite{Born2013-ro}, which applies in the weak scattering regime. For the multiple scattering regime, multi-layer Born approximation model has been developed for 3D phase microscopy \cite{MLB}. Recently, several deep learning-based algorithms have been applied to phase recovery \cite{wang2024use, 10004797}, including CNN \cite{Kamilov:15, Wu:22, Matlock:23}, implicit neural fields \cite{decaf}, and explicit neural fields \cite{FDT}. However, most of these models are designed for transmission geometry under coherent plane wave illumination (except \cite{FDT}), whereas FDT uses partially coherent spherical wave illumination. Furthermore, both image acquisition and reconstruction are time-consuming, as hundreds of images from varying viewpoints must be acquires and processed. To address these challenges, we propose to develop a hybrid representation model with a multi-layer Born approximation to reconstruct the 3D RI from sparse viewpoints.

\section{Method}
\subsection{Prior in Neural Representation}
\subsubsection{Preliminaries}
\textbf{Implicit Neural Representation}\quad Implicit neural representations \cite{nerf} are a type of parametric model that employ a multi-layer perceptron (MLP) to reconstruct 3D properties such as color and density to coordinate systems. These methods have achieved remarkable success across multiple domains, including graphic rendering and 3D scene reconstructions \cite{Takikawa2021Neural,Thies2019Deferred, lu2024fastsparseviewguided}.  Recently, implicit neural representations have also been utilized to enhance microscopic volume reconstructions \cite{decaf, Kang_2024}. 
Here, we represent the estimated RI of sample $\hat{n}$ by $\hat{n}(x,y,z)=F_{inr}(x,y,z)$. where $F_{inr}$ denotes the learnable network. That is, the MLP maps a 3-D spatial coordinate directly to a scalar RI value.

\textbf{Explicit Neural Representation}\quad Explicit neural representations are a non-parametric model first introduced by Plenoxel \cite{Plenoxels}. This approach directly solves object properties at predefined grid points $W_{\mathrm{enr}}$ without an MLP: $\hat{n}(x,y,z)=W_{\mathrm{enr}}(x,y,z).$ This non-parametric strategy is computationally efficient but weakly encodes spatial correlations, hence typically needs more views for high-fidelity results.

\textbf{Triplane Representation}\quad Triplane representation \cite{kplanes_2023, Hexplane} models 3D objects using a low rank grid consisting of three orthogonal feature planes ($W_{xy},W_{xz},W_{yz}$), then fused by a shallow MLP $F_{\mathrm{tri}}$: 
$\hat{n}(x,y,z)=F_{\mathrm{tri}}\bigl(W_{xy}(x,y)+W_{xz}(x,z)+W_{yz}(y,z)\bigr)$. This hybrid design is memory-efficient and excels in surface reconstruction \cite{kplanes_2023, Xu2023DMV3D,Shue20223D,Gupta20233DGen}, though faithfully representing fine volumetric details at the microscale remains challenging.

\subsubsection{Neural Volumetric Prior}
\begin{figure}[htbp]
  \centering
   \includegraphics[width=\linewidth, trim={0 890 550 0}, clip]{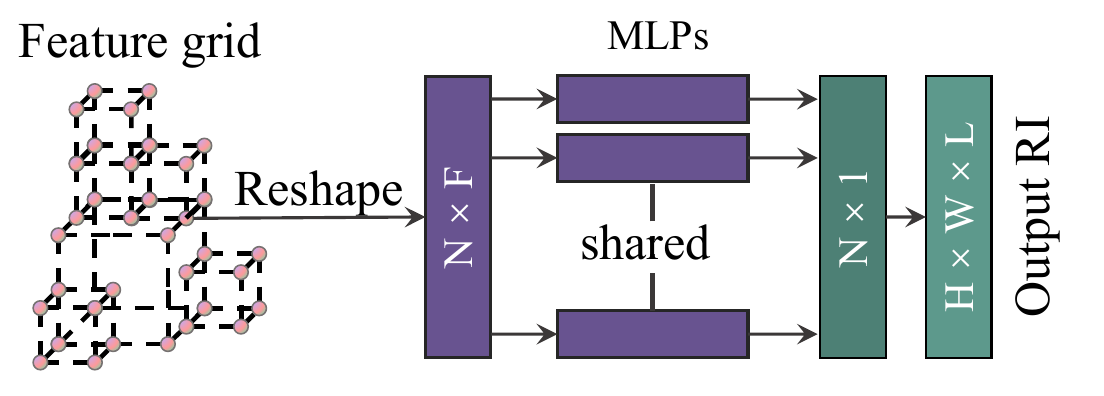}
   \caption{\textbf{NVP architecture.} Adaptive feature grids $(H, W, L, F)$ are first reshaped into feature vectors of size $(N, F)$, where $H, W, L$ are the volume dimensions and $F$ is the feature dimension. These vectors are then passed through MLP layers to predict the RI of each voxel.}
   \label{fig:hybrid}
   \vspace{-4mm}
\end{figure}

To enhance the representation capability, we develop a hybrid model, NVP, that integrates explicit and implicit representations, enabling the network to capture implicit spatial correlations while retaining the sparsity prior of the explicit grid (\cref{fig:overview}a, \cref{fig:hybrid}). The explicit representations store the unknown features of an imaging object in a non-parametric, structural grid, providing direct access to the features at each sampled voxel. The feature grid is adaptive---that is, its resolution dynamically adjusts according to the spatial varying distribution of the RI. Meanwhile, the implicit representations extract dependencies of the features from the explicit neural fields, refining the reconstructed RI and filling in missing values at unmeasured voxels. This hybrid strategy decouples reconstruction quality (i.e., resolution, accuracy) from the number of measurements, effectively overcoming overfitting issues due to sparse views and significantly reducing computational time.

Specifically, we first define a sparse 3D grid for the object with randomly initialized features at each voxel, which are then mapped to corresponding RI by a three-layer MLP (\cref{fig:overview}a, \cref{fig:hybrid}):
\begin{equation}
    \hat{n}(x,y,z)=F_{\mathrm{nvp}}\bigl(W_{xyz}(x,y,z)\bigr),
\end{equation}
where the non-parametric grid \(W_{xyz}(x,y,z)\in\mathbb{R}^{F}\) stores a learnable \(F\)-dimensional feature vector at each voxel, and the MLP \(F_{\mathrm{nvp}}\) converts each vector into a single RI value.  

The output of the MLP is a volumetric object with estimated RI, which becomes the input to the physics-informed rendering equation, along with self-calibrated fluorescent illumination (\cref{fig:overview}b). The loss function is calculated between the predicted and measured images (\cref{fig:overview}c). Details of the rendering equation and loss function are presented in the following subsections.

\subsection{Physical prior in the Rendering Equation using Diffraction Optics}\label{Sect 3.3}
\noindent\textbf{How is physically accurate 3D RI reconstruction of microscale objects achieved from sparse views?}\quad 

We model the rendering process using a multi-slice Born approximation that accounts for light diffraction at the microscale \cite{Tian:15, Chowdhury:19, Chen:20, MLB, Xue:22}. In diffractive optics, when light propagates through a transparent sample with spatially varying RI (Figure \ref{fig:overview}b), each voxel alters the phase of the wavefront. As light continues to propagate, these phase modulations accumulate and interfere, resulting in diffraction patterns that carry information about the 3D RI distribution. The rendering equation describes this process by modeling the imaging volume as a stack of $N_z$ thin slices with spatially varying RI. The light field $\hat{E}_{k,i}(\mathbf{r})$ at $k^{th}$ slice illuminated by the $i^{th}$ fluorescence source is defined as:

\begin{equation}
\hat{E}_{k,i}(\mathbf{r}) = \mathcal{P}_{\Delta z} \left\{ t_k(\mathbf{r}) \cdot \hat{E}_{k-1,i}(\mathbf{r}) \right\},
\label{diffraction}
\end{equation}
where $\mathcal{P}_{\Delta z}$ denotes the propagation operator, and $t_k(\mathbf{r})$ is the transmission function of the $k^{th}$ slice and is related to its RI. After transmitting through the final slice, the accumulated light field $\hat{E}_{N_z}(\mathbf{r})$ is captured by a camera as an intensity-only image $\hat{I}_i(\mathbf{r})$:
\begin{equation}
    \hat{I}_i(\mathbf{r}) = \left| \hat{E}_{N_z,i}(\mathbf{r}) \right|^2.
\end{equation}
The intensity images under illumination from each fluorescence source $\hat{I}_1, \hat{I}_2, ..., \hat{I}_n$ are computed using this procedure, which are the sparse views. Further details on the mathematical formulation are provided in the Supplementary~1.

We implemented this diffraction-based multi-slice rendering equation on a GPU (see details in Section 4), achieving fast rendering through parallel computing and predefined propagation kernels. This rendering equation reforms conventional ray optics by introducing a more accurate, physics-informed prior while maintaining comparable computational efficiency. 

\subsection{Essential Module for Processing Experimentally Captured Data}\label{Sect 3.4}
\textbf{Coherent Alignment}
The diffraction of coherent light in the rendering equation is different from the partially coherent or incoherent light (e.g., fluorescence) used in the experiment. To mitigate the model mismatching, we incorporate a coherent mask in the model that accurately and efficiently aligns measured fluorescence images with the model-predicted images. More details are provided in Supplementary~2.

\begin{figure*}[htbp]
    \centering
    \includegraphics[width=0.99\linewidth, trim={0 340 0 0}, clip]{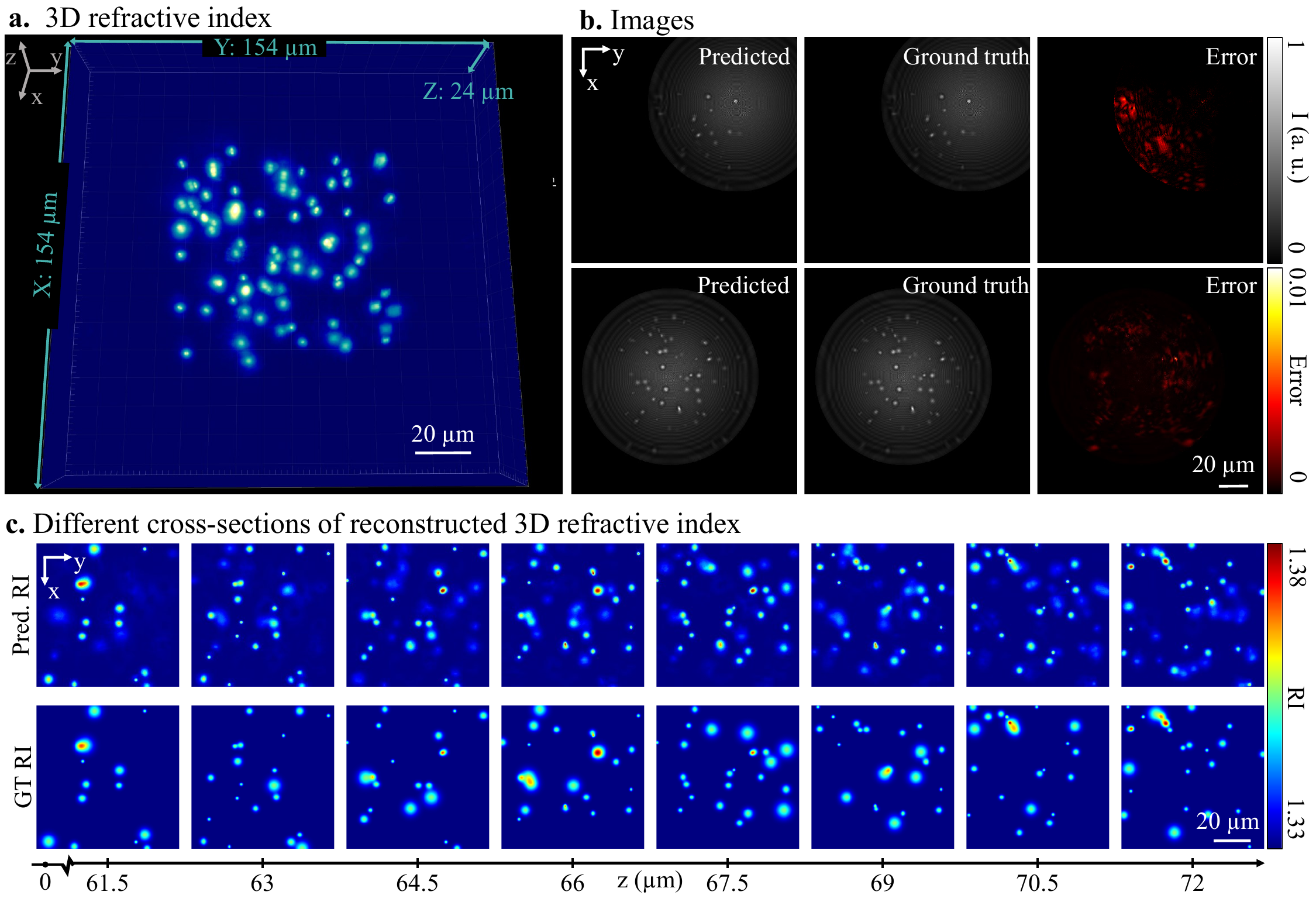}
   \caption{
    \textbf{NVP reconstructs the 3D RI of the synthetic cell dataset with micron-scale resolution and excellent optical sectioning.}
    (a) Reconstructed 3D RI distribution showing the spatial distribution of cells within a 154 $\mu$m $\times$ 154 $\mu$m $\times$ 24 $\mu$m volume. 
    (b) Predicted fluorescence images under the illumination from two representative fluorescence sources, demonstrating the accuracy of the reconstructed model compared to the ground-truth data.
    (c) Cross-sectional comparison of RI reconstructed by NVP (top) and ground-truth (bottom) along the axial direction, illustrating consistency between the reconstructed and ground-truth RI distribution across the sample volume. Color bars indicate intensity ranges for predicted images, absolute errors, and RI values.
    }
   \label{fig:beads}
   \vspace{-4mm}
\end{figure*}

\textbf{Self-Calibration of Viewpoints}
Calibrating viewpoints is crucial \cite{sparf,martin2021nerf,Barf} for 3D rendering with the multi-slice model, as the quality and accuracy of the 3D reconstruction depend on precise localization of the illumination sources. 
We propose a self-supervised calibration method that estimates fluorescence source positions from the fluorescence images by Gaussian fitting and jointly optimizes them with the MLP parameters to improve 3D reconstruction quality. We have conducted ablation study on the self-calibration of viewpoints. More details are provided in Supplementary~7 and Table \ref{tab:noise_fixdx}.

\section{Experiments}\label{Sect4}
\subsection{Experimental Settings}\label{Sec 4.1}
\noindent\textbf{Computational setting}\quad 
We train our model on an NVIDIA A6000 GPU with 48 GB memory using the PyTorch framework. The Adam optimizer is used with an initial learning rate of \(5 \times 10^{-3}\), a momentum decay rate of 0.9, and a squared gradient decay rate of 0.99. The batch size is set to 10. For the results presented in this paper, the typical computation time for 3D RI reconstruction is around 20 minutes. More details on the implementation are provided in Supplementary 6.

\noindent\textbf{Image Acquisition}\quad  
We implement the optical setup as described in \cite{FDT, Xue2019, Li2024-sp}. The imaging objects are label-free cultured cells in a petri dish coated with fluorescent paint on the outer surface. Fluorescence from a single spot is excited by focusing a laser beam, which is modeled as a point source. The excitation positions are randomly selected. As the fluorescence propagated through the cells, it is diffracted and captured by a high-sensitivity sCMOS camera (Kinetix22).

\noindent\textbf{Loss and Metrics Design}\quad 
We formulate the loss function to minimize the difference between the predicted and ground-truth images, while applying a regularization term on the predicted 3D RI to promote smoothness and preserve fine structural details. The overall loss is expressed as $
    \mathcal{L} = \mathcal{L}_{\text{img}} + \tau \mathcal{R}_{\text{ri}},
$ where $\mathcal{L}_{\text{img}}$ combines multiple image-level loss terms such as $L_1,~L_2,~SSIM$, and $\mathcal{R}_{\text{ri}}$ is the loss associated with total-variation regularization (Supplementary~4).


\begin{figure*}[htbp]
  \centering
   \includegraphics[width=0.97\linewidth, trim={0 700 0 0}, clip]{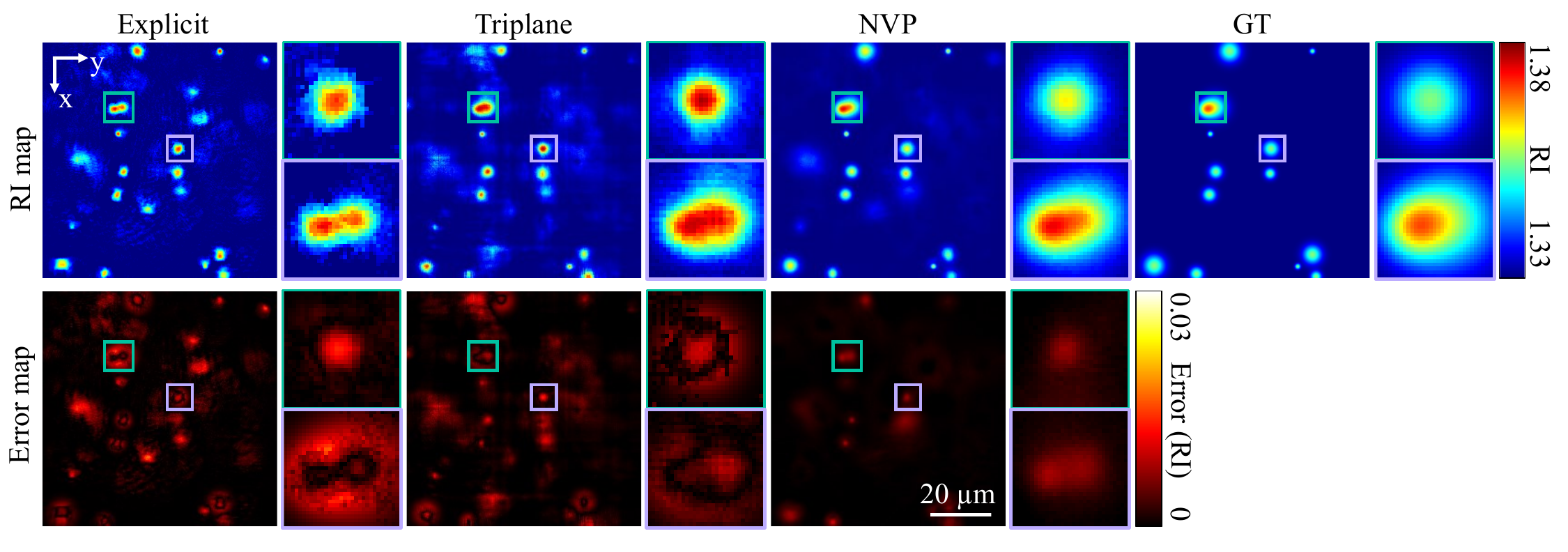}
    \caption{\textbf{NVP achieves more accurate reconstruction compared to the explicit and triplane methods}. The top row shows the RI maps produced by each method, with the ground-truth RI map included for reference. The bottom row displays the corresponding error maps, highlighting the reconstruction errors in each method. NVP demonstrates superior performance, producing smoother reconstructions and lower errors that closely match the ground truth. }
   \label{fig:beads_comp}
   
\end{figure*}

\begin{table*}[ht]
\centering
    \begin{minipage}{0.65\linewidth}
        \centering
        
        \resizebox{1\linewidth}{!}{ 
        \begin{tabular}{l c ccc ccc ccc}
        \toprule
         \multicolumn{2}{c}{Synthetic}& \multicolumn{3}{c}{6 illuminations} & \multicolumn{3}{c}{7 illuminations} & \multicolumn{3}{c}{20 illuminations} \\
         \multicolumn{2}{c}{cells} 
         & PSNR $\uparrow$ & SSIM $\uparrow$ & LPIPS $\downarrow$ 
         & PSNR $\uparrow$ & SSIM $\uparrow$ & LPIPS $\downarrow$ 
         & PSNR $\uparrow$ & SSIM $\uparrow$ & LPIPS $\downarrow$ \\
        \midrule
        \multirow{2}{*}{Exp} 
        & RI  & 28.6203 & \cellcolor[HTML]{ECE7FF}0.8540 & \cellcolor[HTML]{ECE7FF}0.1821 & 28.7268 & \cellcolor[HTML]{ECE7FF}0.8469 & 0.1777 & 28.8803 & 0.8650 & 0.1188 \\
        & IMG & 58.2588 & 0.9999 & 0.0001 & 57.3599 & 0.9999 & 0.0002 & 55.4224 & 0.9996 & 0.0003 \\
        \multirow{2}{*}{Tri} 
        & RI  & \cellcolor[HTML]{ECE7FF}28.7288 & 0.7128 & 0.2152 & \cellcolor[HTML]{ECE7FF}30.4342 & 0.7619 & \cellcolor[HTML]{ECE7FF}0.1385 & \cellcolor[HTML]{ECE7FF}30.6149 & \cellcolor[HTML]{E0D9FF}0.9621 & 0.0620 \\
        & IMG & 55.4019 & 0.9997 & 0.0003 & 56.1654 & 0.9997 & 0.0002 & 47.1665 & 0.9970 & 0.0028 \\
        \multirow{2}{*}{Ours} 
        & RI  & \cellcolor[HTML]{E0D9FF}30.7318 & \cellcolor[HTML]{E0D9FF}{0.8910} & \cellcolor[HTML]{E0D9FF}0.1034 & \cellcolor[HTML]{E0D9FF}30.9643 & \cellcolor[HTML]{E0D9FF}{0.8966} & \cellcolor[HTML]{E0D9FF}0.0897 & \cellcolor[HTML]{E0D9FF}31.3787 & \cellcolor[HTML]{ECE7FF}{0.8959} & \cellcolor[HTML]{E0D9FF}0.0543 \\
        & IMG & 71.0492 & 1.0000 & 0.0000 & 69.9748 & 1.0000 & 0.0000 & 68.6298 & 1.0000 & 0.0000 \\
        \bottomrule
        \end{tabular}
        }
        \caption{Reconstruction quality metrics for synthetic cells dataset across different illuminations and various methods in RI and predicted image results.}
        \label{tab:cell}
    \end{minipage}%
    \hspace{0.01\linewidth}
    \begin{minipage}{0.278\linewidth}
        \centering
        \resizebox{1\linewidth}{!}{ 
        \begin{tabular}{l c ccc}
        \toprule
        \multicolumn{2}{c}{Synthetic} & \multicolumn{3}{c}{7 illuminations}\\
        \multicolumn{2}{c}{tissue} & PSNR $\uparrow$ & SSIM $\uparrow$ & LPIPS $\downarrow$ \\
        \midrule
        \multirow{2}{*}{Exp} 
        & RI  & \cellcolor[HTML]{ECE7FF}11.6144 & \cellcolor[HTML]{ECE7FF}0.2954 & \cellcolor[HTML]{ECE7FF}0.5676 \\
        & IMG & 35.7319 & 0.9168 & 0.0329 \\
        \multirow{2}{*}{Tri} 
        & RI  & 9.0632 & 0.1323 & 0.7122 \\
        & IMG & 46.4842 & 0.9956 & 0.0040 \\
        \multirow{2}{*}{NVP} 
        & RI  & \cellcolor[HTML]{E0D9FF}11.6412 & \cellcolor[HTML]{E0D9FF}{0.4775} & \cellcolor[HTML]{E0D9FF}0.5038 \\
        & IMG & 62.7914 & 0.9999 & 0.0001 \\
        \bottomrule
        \end{tabular}

        }
        \caption{Reconstruction quality metrics for synthetic tissue.}
        \label{tab:tissue}
    \end{minipage}
    \vspace{-4mm}
\end{table*}

\subsection{Data Preparation}
We evaluate our method on two in-house synthetic 3D datasets: synthetic cells and synthetic tissue. The synthetic cells are designed to assess the resolution across varying RI and the optical sectioning ability across different axial planes (Fig. \ref{fig:beads}). Each synthetic cell is modeled as a 3D Gaussian function $\mathcal{N}(\mathbf{\mu}, \sigma)$, where $\mathbf{\mu}$ denotes its 3D position and $\sigma$ denotes its radius ($\sigma = 5-10$ pixels). The synthetic cells are randomly distributed within a volume of $512 \times 512 \times 12$ voxels with a voxel size of $\{0.3, 0.3, 2\} \, \mu m$ along the ${x, y, z}$ axes. The synthetic tissue dataset consists of blood vessels and neurons spanning a 3D volume and is designed to assess reconstruction performance across different morphological structures (spatial frequencies). The 3D volume consists of $512 \times 512 \times 15$ voxels, with a voxel size of $\{0.9, 0.9, 3\} \, \mu m$. In both datasets, 100 ground-truth fluorescence images ($512 \times 512$ pixels) are generated by the rendering equation with the illumination of 100 distinct fluorescence sources. It is worth noting that NVP typically uses only 6-7 fluorescence images randomly selected from this set. 

\begin{figure*}[htbp]
  \centering
   \includegraphics[width=0.99\linewidth, trim={0 550 0 0}, clip]{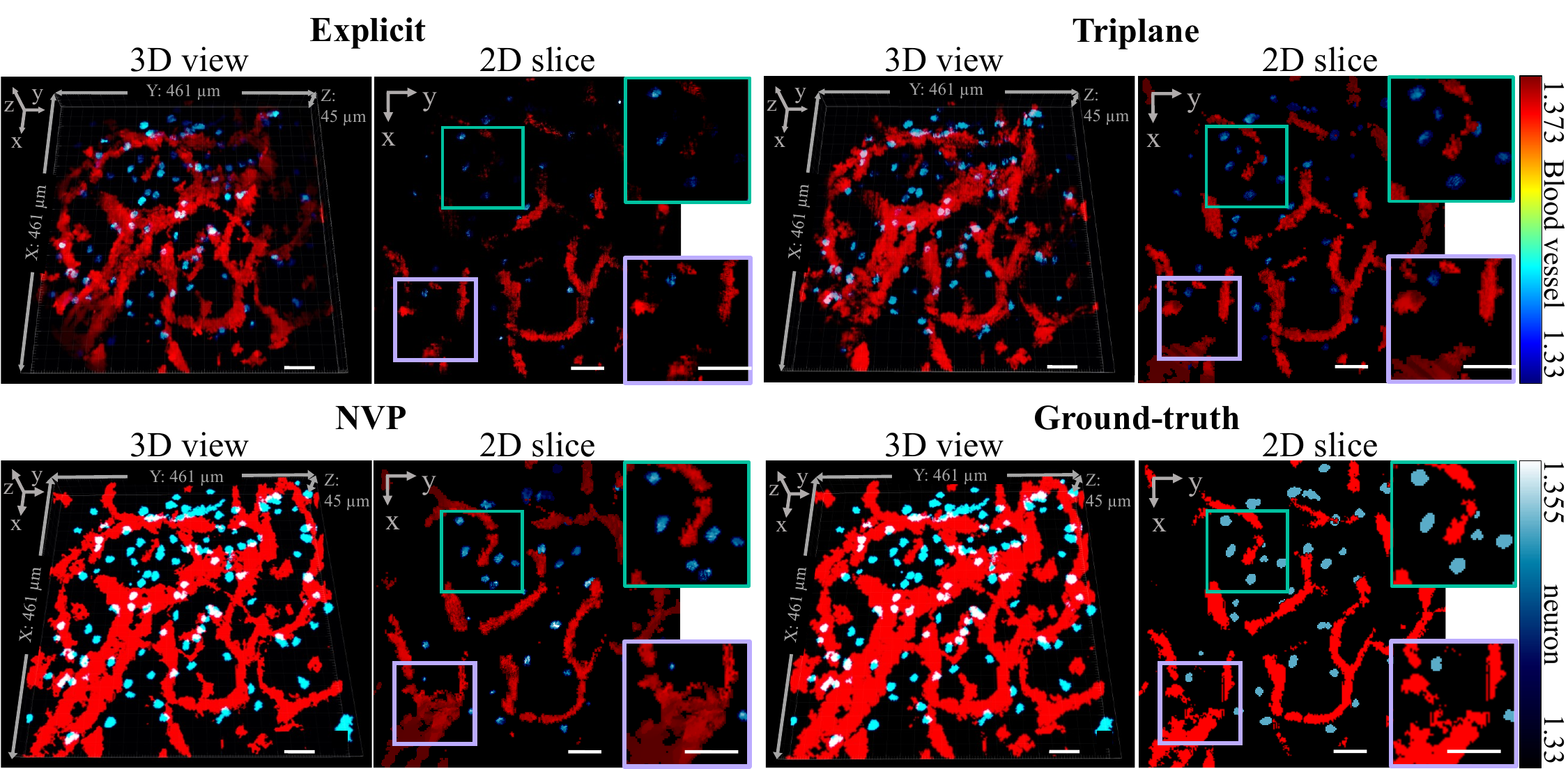}
    \caption{
    \textbf{NVP reconstructs the RI of varying morphological structures in the synthetic tissue dataset and outperforms the explicit and triplane methods.} The synthetic tissue contains both continuous structures (blood vessels, red) and sparse structures (neurons, blue). Left: the 3D view of the reconstructed RI. Right: a representative $Z$-slice and zoomed-in views of the reconstructed RI. Scale bars denote 50 $\mu$m.
    }
    \label{fig:neuron} 
    \vspace{-5mm}
\end{figure*}

\subsection{Comparison of Neural Representation Methods}

\noindent\textbf{Baseline}\quad Since our task is to reconstruct 3D RI from microscopic images using a diffraction-based optical model, few existing methods are available for direct comparison. We implement the explicit approach \cite{FDT} as a baseline in our experiment, as it is the only closely related method. Furthermore, we adapt the triplane approach to this task for comparison.

\noindent\textbf{Quantitative Reconstruction Results}\quad 
We first assess the resolution and optical sectioning capability of NVP by reconstructing the 3D RI of synthetic cells (Figure \ref{fig:beads}). The synthetic cells, with sizes ranging from 3.53 $\mu m$ to 7.05 $\mu m$ (full-width-half-maximum), are randomly distributed within a 3D volume. The NVP model is configured with a feature dimension of 16 and 6 network layers (see details in Supplementary Section 6, Table \ref{tab:feature}, and Table \ref{tab:layer}). In the reconstructed volume (Figure \ref{fig:beads}a), all cells are clearly distinguishable from one another and from the background, indicating the high-resolution reconstruction of RI. The predicted fluorescence images also closely match the ground-truth images (Figure \ref{fig:beads}b). Furthermore, the reconstructed RI on each $Z$-plane is nearly identical to the ground-truth RI at the corresponding plane, indicating excellent optical sectioning performance (see Supplementary Section 3 for details). These results demonstrate that NVP accurately reconstructs high-resolution 3D RI.

Next, we quantitatively compare the reconstruction results of the synthetic cells using different neural representation methods with the ground-truth RI (Figure \ref{fig:beads_comp}). The RI reconstructed by NVP is more accurate and contains fewer artifacts compared to the results from the explicit and triplane methods, especially in fine structures. The error maps further support this conclusion, showing that NVP yields notably lower errors, particularly in dense or complex regions, indicating its ability to preserve fine details while minimizing noise.

To evaluate NVP's reconstruction performance across different morphological structures, we reconstruct the RI of synthetic tissue using NVP and compare the results with those from the explicit and triplane methods, as well as the ground truth (Figure \ref{fig:neuron}). The explicit method exhibits structural artifacts and lacks detail in complex areas, while the triplane method reduces some artifacts but introduces grid effects, particularly in dense regions. In contrast, NVP closely matches the ground truth, achieving a significantly higher SSIM of 0.4775 for RI, compared to 0.2954 for explicit method and 0.1323 for triplane method  (Table \ref{tab:tissue}). NVP preserves fine details and minimizes artifacts in both sparse and continuous structures, demonstrating superior robustness and accuracy. These results highlight NVP’s effectiveness for high-fidelity reconstruction.



\subsection{Effect of Varying Number of Views}

To evaluate reconstruction from sparse views, we reconstruct the RI of synthetic cells using 6, 7, and 20 fluorescence images with NVP, the explicit method, and the triplane method. We quantitatively evaluate the results by comparing PSNR, SSIM \cite{ssim}, and LPIPS \cite{lpips} for both the RI and the predicted images (Table \ref{tab:cell}). With as few as 6 fluorescence images, NVP outperforms the other methods in RI reconstruction, achieving a PSNR of 30.73, SSIM of 0.8910, and LPIPS of 0.1034, demonstrating strong structural fidelity and perceptual quality. When the number of fluorescence images is reduced from 20 to 6, the performance of NVP shows minimal performance degradation, with only a 0.65 decrease in PSNR and a 0.005 decrease in SSIM. Furthermore, NVP with 6 images achieves results comparable to explicit and triplane methods using 20 illuminations. For predicted images, NVP consistently achieves near-perfect scores, with SSIM reaching 1.0000 and LPIPS approaching zero across all cases, indicating excellent alignment with the ground truth. We also evaluate the PSNR, SSIM, and LPIPS for reconstructed synthetic tissue with these methods and observe similar conclusions (Table \ref{tab:tissue}, Table \ref{tab:tissue2}). More details are provided in the Supplementary 5. These results highlight NVP’s robustness to sparse views, making it highly suitable for applications with limited data acquisition, such as imaging cardiomyocyte contractions and embryo development. The ability to achieve high-quality 3D reconstruction with fewer 2D images enables faster imaging and reduces photodamage to biological specimens.

\begin{figure*}[htbp]
  \centering
   \includegraphics[width=0.99\linewidth, trim={0 670 40 0}, clip]{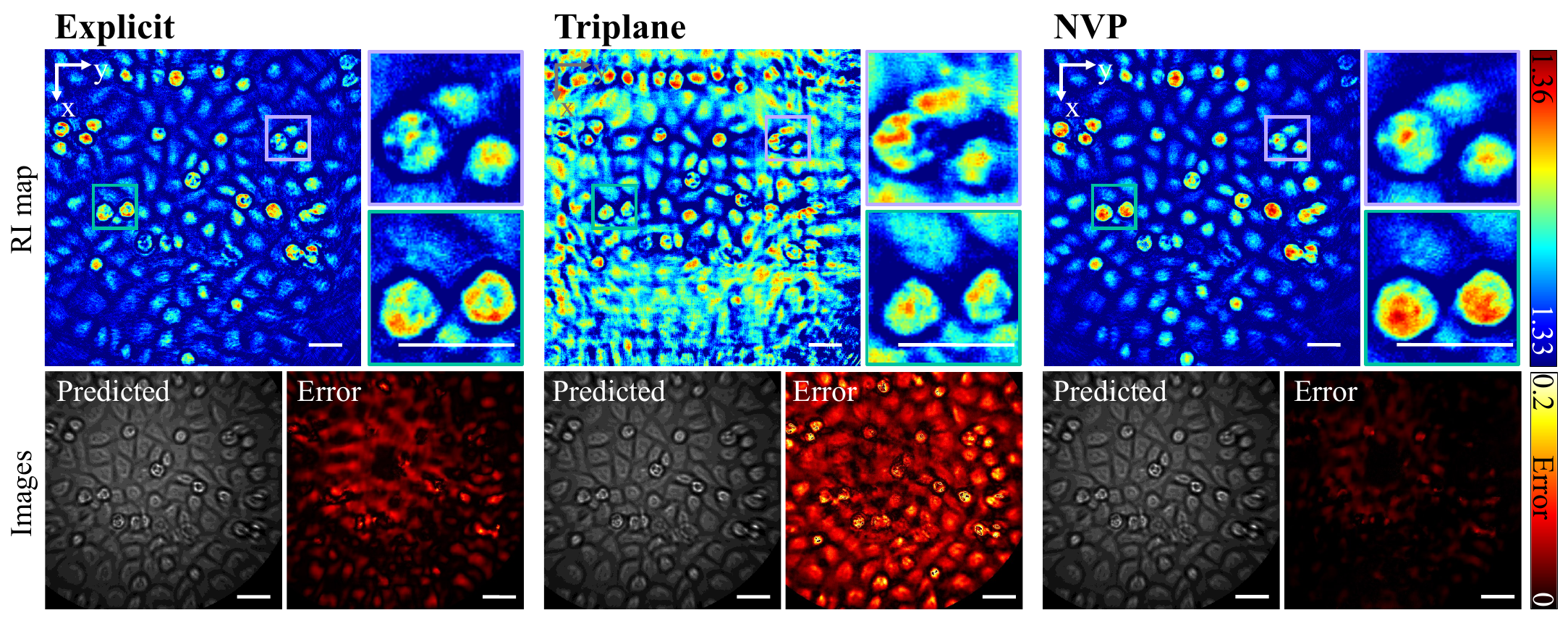}
    \caption{\textbf{NVP reconstructs the 3D RI of living MDCK cells with superior accuracy and fewer artifacts compared to the results from the explicit and triplane methods.} 
    RI maps (top row), predicted images (bottom row, left), and error maps (bottom row, right) are shown for the explicit method, the triplane method, and NVP. The zoomed-in views (purple and green boxes) of the RI maps are shown in the top row, right. Scale bars denote 20 $\mu$m.
    }
    
    \label{fig:mdck}
    \vspace{-5mm}
\end{figure*}

\subsection{Experiments on Real Biological Samples}

We apply the explicit, triplane, and NVP methods to reconstruct the 3D RI of living MDCK cells from 19 fluorescence images captured with our custom-built microscope system described in Section \ref{Sec 4.1} (Figure \ref{fig:mdck}, Table \ref{tab:metrics2}). Among the three methods, NVP reconstructs cellular structures more accurately than the explicit and triplane methods, since it uses the full-dimensional representatives informed by prior knowledge of features. The RI reconstructed by the explicit method shows discontinuities and noise due to absence of prior knowledge of data distribution, making it more prone to local minima. Meanwhile, the triplane method displays severe grid artifacts, as its three-plane decomposition limits its ability to reconstruct high-dimensional representation. This limitation has been observed in previous applications, such as MRI volume reconstruction \cite{triplane_MR}. The predicted images and error maps further confirm that NVP surpasses both baselines in the accuracy of predicted images. Quantitative metrics show that NVP achieves the highest SSIM (0.9977) and PSNR (40.7043), and the lowest LPIPS (0.0015), indicating superior structural accuracy and perceptual fidelity compared to the other methods (Table \ref{tab:metrics2}). While the explicit method slightly outperforms the triplane method in structural metrics, it has higher overall errors. In summary, these results demonstrate that NVP is able to quantitatively reconstruct the 3D RI of real biological samples from limited fluorescence images with high accuracy and robustness. At the same resolution, our method converges in 20 minutes to SSIM$>0.99$, whereas the explicit baseline \cite{FDT} requires 60 minutes.

\begin{table}[ht]
    \centering
    \begin{tabular}{cccc}
    \toprule
    \textbf{Methods}  & \textbf{SSIM}$\uparrow$ & \textbf{LPIPS}$\downarrow$ & 	\textbf{PSNR}$\uparrow$  \\ 
    \midrule
    \multirow{1}{*}{Explicit}  & \cellcolor[HTML]{ECE7FF}0.9944 & \cellcolor[HTML]{ECE7FF}0.0051 & \cellcolor[HTML]{ECE7FF}36.6540 \\ 
    \multirow{1}{*}{Triplane}  & 0.9762 & 0.0285 & 32.6921 \\ 
    \multirow{1}{*}{NVP}  & \cellcolor[HTML]{E0D9FF}0.9977 & \cellcolor[HTML]{E0D9FF}0.0015 & \cellcolor[HTML]{E0D9FF}40.7043 \\ 
    \bottomrule
    \end{tabular}
    \caption{Comparison of NVP, explicit, and triplane methods based on the quality of predicted images of MDCK cells.}
    \label{tab:metrics2}
    \vspace{-5mm}
\end{table}
\section{Conclusion and Discussion}\label{discussion}
\vspace{-2mm}

We present a new neural representation method, NVP, for efficient and accurate volumetric RI representation from limited and sparse-view microscopic images acquired using FDT. NVP uses hybrid representations by integrating the explicit adaptive feature grid with nonlinear MLP to enhance the quality and efficiency of 3D reconstruction of the RI spatial distribution. NVP also incorporates a rendering equation based on diffraction optics, which accurately models phase changes as light interacts with biological samples. This enables high-fidelity, high resolution rendering of microscale structures and quantitative volumetric reconstruction, while maintaining comparable processing time. To address model mismatches when processing experimentally captured images, we incorporate coherent alignment and self-calibration. We validate NVP using experiments on both synthesized data and real biological samples. NVP demonstrates superior performance over the explicit and triplane methods, achieving a 50-fold reduction in the number of required measurements and a 3-fold increase in processing speed compared to previous work \cite{FDT}. These results highlight NVP's potential for high-fidelity, high-speed volumetric reconstruction using FDT.

Compared to the explicit method, NVP introduces a neural prior that is significantly more effective than the sparse prior used in the explicit method. Compared to the triplane method, NVP retains the full representation dimensions, allowing it to better preserve high-dimensional representations and produce fewer artifacts. NVP can be extended to other imaging modalities and datasets by adjusting the underlying physics model. Its performance may be further improved by integrating 3D convolution kernels to capture spatial relationships and off-grid features, or leveraging data priors such as masked auto-encoders \cite{nerfvae} or diffusion models \cite{nerfdiff, chen2023singlestagediffusionnerfunified}.

\section{Acknowledgement}\label{Sect6}

We thank Dr. Sara Fridovich-Keil at Georgia Institute of Technology for insightful discussions on representation methods. We thank Ziqiao Ma at the University of Michigan for proofreading the manuscript. Research reported in this publication was supported by the National Institute of General Medical Sciences of the National Institutes of Health 1R35GM155193-01 and the National Science Foundation CAREER Award 2443604 to Dr. Yi Xue. This work was also supported by Dr. Yi Xue’s startup funds from the Department of Biomedical Engineering at the University of California, Davis.
{
    \small
    \bibliographystyle{unsrt}
    \bibliography{main}
}
\clearpage
\setcounter{page}{1}
\maketitlesupplementary
\setcounter{section}{0}

\section*{Overview}
This supplementary material consists of 7 sections that provide extended data and insights into the Neural Volumetric Prior (NVP). First, detailed equations describing the rendering process based on diffraction optics are provided. The second section elaborates on the coherent alignment for processing experimentally captured images. The third section discusses the optical section of FDT using k-space analysis. The fourth section specifies the loss functions. The fifth section explains evaluation metrics used to design and optimize NVP. The sixth section provides implementation details, including the optimization of feature dimensions and network layers. The seventh section presents an ablation study on self-calibration. 

Furthermore, four supplementary tables are included to present extended results. Table~\ref{tab:tissue2} compares the reconstruction performance of various methods across different numbers of measurements, extending Table~\ref{tab:tissue}. Table~\ref{tab:feature} examines the effect of feature dimensionality on NVP’s performance. Table~\ref{tab:layer} evaluates the impact of network depth on NVP’s performance. Table~\ref{tab:noise_fixdx} provides an ablation study investigating the role of self-calibration for experimentally captured data.

\section{Multi-slice Model for Rendering Process}\label{multislice}

To accurately model light propagation through a 3D heterogeneous semi-transparent sample, the sample is represented by multiple slices overlapping in the light propagation direction, so called the "multi-slice model". To improve computational efficiency, we extend the conventional multi-slice model from classical optimization algorithms (i.e., FISTA) \cite{Xue:22} to the PyTorch framework. This differentiable model allows automated backpropagation. It also incorporates a coarse-to-fine strategy for adjusting parameters and fine-tuning the RI. 

In this model, the 3D phase object is represented as a stack of $N_z$ layers, each with an unknown RI, $\hat{n}_k(\mathbf{r})$ for \( k = 1, 2, ..., N_z \), where:

\begin{equation}
\hat{n} \triangleq \left\{ \hat{n}_k(\mathbf{r}) \right\}_{k=1}^{N_z}, \quad \mathbf{r} = (x, y),
\end{equation}

As light propagates through each layer, its phase is altered according to the transmission function \( t_k(\mathbf{r}) \):

\begin{equation}
t_k(\mathbf{r}) = \exp \left( \frac{j 2\pi}{\lambda} \Delta z (\hat{n}_k(\mathbf{r}) - n_b) \right),
\end{equation}

where \(\lambda\) is the wavelength, \(\Delta z\) is the layer thickness, and \(n_b\) is the background RI. The Fresnel propagation operator \(\mathcal{P}_{\Delta z}\) represents light propagation through the layers:

\begin{equation}
\mathcal{P}_{\Delta z} \{ \cdot \} = \mathcal{F}^{-1} \left\{ \exp\left(-j 2\pi \Delta z \sqrt{\left(\frac{1}{\lambda}\right)^2 - ||\mathbf{k}||^2} \right) \cdot \mathcal{F} \{ \cdot \} \right\},
\end{equation}

where \(\mathcal{F}\{\cdot\}\) and \(\mathcal{F}^{-1}\{\cdot\}\) are the Fourier transform and its inverse. The electric field \(\hat{E}_{k,i}(\mathbf{r})\) propagating through the $k$-th layer for the $i$-th fluorescent source is:

\begin{equation}
\hat{E}_{k,i}(\mathbf{r}) = \mathcal{P}_{\Delta z} \left\{ t_k(\mathbf{r}) \cdot \hat{E}_{k-1,i}(\mathbf{r}) \right\}.
\end{equation}

At the image plane, the electric field is calculated by applying the pupil function \( p(\mathbf{k}) \):

\begin{equation}
\hat{E}_i(\mathbf{r}) = \mathcal{F}^{-1} \left\{ p(\mathbf{k}) \cdot \mathcal{F} \left\{ \hat{E}_{N_z,i}(\mathbf{r}) \right\} \right\}.
\end{equation}

For thick objects, the field is back-propagated over a distance \(\Delta Z_c\) before reaching the camera:

\begin{equation}
\hat{E}_i(\mathbf{r}) = \mathcal{F}^{-1} \left\{ p(\mathbf{k}) \cdot \mathcal{F} \left\{ \mathcal{P}_{-\Delta Z_c} \left\{\hat{E}_{N_z,i}(\mathbf{r}) \right\} \right\}\right\}.
\end{equation}

The camera detects the intensity of the light field as:

\begin{equation}
\hat{I}_i(\mathbf{r}) = | \hat{E}_i(\mathbf{r}) |^2.
\end{equation}

\section{Coherent Alignment}\label{Coherent}

\begin{figure}[htbp]
  \centering
   \includegraphics[width=0.99\linewidth, trim={0 740 600 0}, clip]{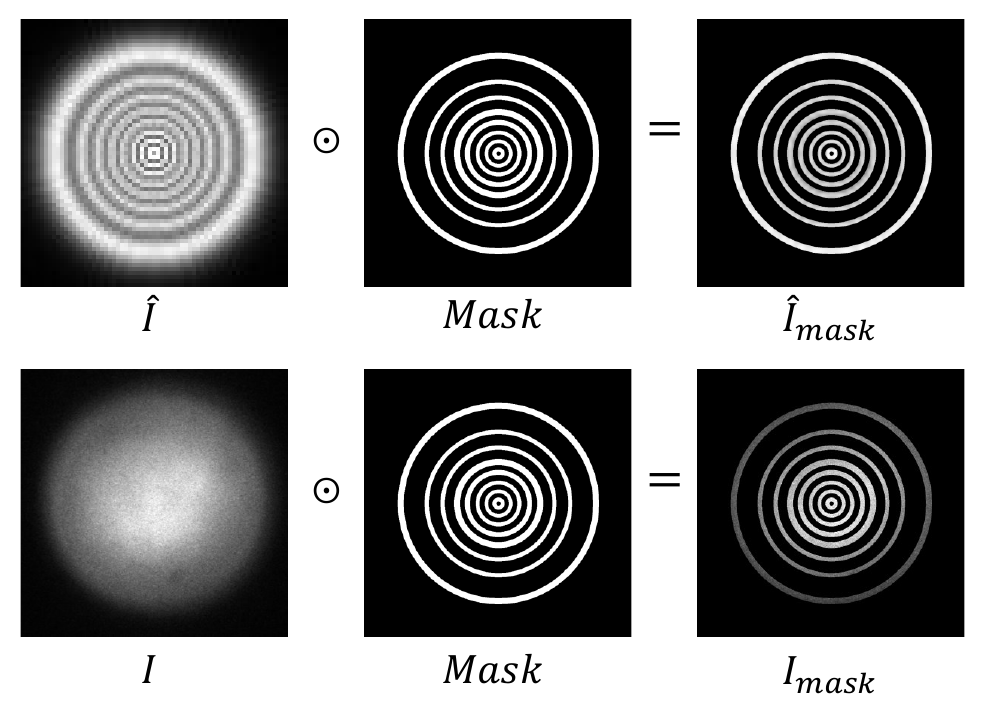}
    \caption{\textbf{Illustration of the coherent alignment method to address partial coherence in fluorescence-based illumination.} The top row shows the synthetic image \(\hat{I}\) (left) multiplied by the coherent mask (center) to produce the masked synthetic image \(\hat{I}_{\text{mask}}\) (right). The bottom row depicts the real image \(I\) (left) undergoing the same masking process to generate the masked real image \(I_{\text{mask}}\) (right). This approach aligns the partially coherent fluorescence illumination with the coherent multi-slice model, mitigating mismatches caused by variations in photon frequency and phase.}
    \label{fig:mask}
\end{figure}

The multi-slice model allows us to generate synthetic images using rendering equations under the assumption that the illumination source is coherent. Coherence implies that all photons emitted from the source share the same frequency and phase. However, in the experimental data captured by FDT, the light source (i.e., fluorescence) is partially coherent, introducing model mismatching.

One approach to address this mismatch would be to mathematically model \cite{bastiaans1986application} the partially coherent sources to generate more accurate synthetic images. However, this approach is impractical due to the inherent randomness and spatially-variant distribution of fluorescence light. 

To overcome this limitation, we propose a coherent alignment method (Figure~\ref{fig:mask}). The synthetic image \(\hat{I}\) (top left), generated using coherent light sources, is adjusted to match partial coherence by applying a coherent mask derived from the diffraction pattern (top center). This results in the masked synthetic image \(\hat{I}_{\text{mask}}\) (top right). A similar process is applied to the real observed image \(I\) (bottom left), which undergoes masking with the same coherent mask (bottom center) to produce the masked real image \(I_{\text{mask}}\) (bottom right). The loss is calculated between \(I_{\text{mask}}\) and \(\hat{I}_{\text{mask}}\) instead of $I$ and \(\hat{I}\), which reduces model mismatches. 

\section{Missing cone problem.} 
\begin{figure}[htbp]
\vspace{-3mm}
    \centering
    \includegraphics[width=0.99\linewidth, trim={0 1000 630 00}, clip]{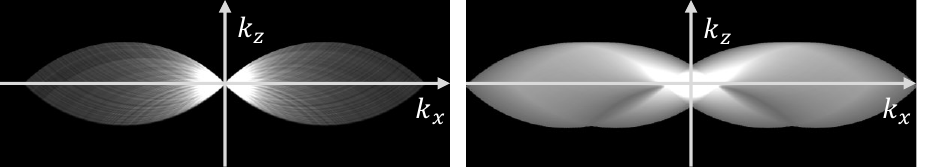}
   \caption{
    \textbf{OTF under coherent (left) and partially coherent (right) conditions.}}
   \label{fig:otf}
   \vspace{-3mm}
\end{figure}

NVP solves the missing cone problem by using fluorescence as a partially coherent light source for illumination. It is well established that partially coherent light can solve this issue \cite{WAX20022256}. We also provide the optical transfer function (OTF) of partially coherent illumination (Figure ~\ref{fig:otf}), which indicates that it fills the missing cone compared to traditional coherent illumination, thereby enabling optical sectioning for 3D reconstruction.

\section{Loss}\label{loss}

The image reconstruction loss $\mathcal{L}_{\text{img}}$ is a weighted sum of $L_1$, $L_2$, and Structural Similarity Index (SSIM) losses, ensuring robustness to pixel-wise differences while preserving perceptual quality. This loss is defined as:
\begin{equation}
    \mathcal{L}_{\text{img}} = \frac{1}{N} \sum_{i=1}^{N} \alpha \| \hat{I}_i - I_i \|_1 + \beta \| \hat{I}_i - I_i \|_2^2 + \gamma \, \mathscr{\ell}_{\text{SSIM}}(\hat{I}_i, I_i),
\end{equation}
where $N$ is the number of images, $\hat{I}_i$ and $I_i$ are the predicted and ground-truth images, and $\alpha$, $\beta$, and $\gamma$ are the corresponding weights for each loss term.

Since the RI distribution in biological samples varies gradually, total variation regularization is applied across both the $X$-$Y$ plane and the $Z$-axis to enforce smoothness in the predicted 3D RI. This is formulated as:
\begin{equation}
    \mathcal{R}_{\text{ri}} = \lambda_{xy} \, TV_{xy}(\hat{n}) + \lambda_z \, TV_z(\hat{n}),
\end{equation}
where $TV_{xy}(\hat{n})$ and $TV_z(\hat{n})$ represent the total variation for predicted RI $\hat{n}$ along the $X$-$Y$ plane and $Z$-axis, respectively. The weights $\lambda_{xy}$ and $\lambda_z$ control the strength of regularization in each direction, the $\alpha, \beta, \gamma={4,4,1.5}$..

\section{Metrics}\label{Metrics}

We use three metrics to evaluate the quality of our 3D reconstructions: SSIM, Learned Perceptual Image Patch Similarity (LPIPS), and Peak Signal-to-Noise Ratio (PSNR). These metrics provide a comprehensive evaluation: SSIM for structural similarity, LPIPS for perceptual quality, and PSNR for signal fidelity. For evaluation, we normalize te image intensities to a range of $0-1$, while preserving the original reconstructed RI values, as they retain their real-world physical significance. 

\subsection{SSIM}
SSIM measures the similarity between two images in terms of luminance, contrast, and structure, with values ranging from 0 to 1 (higher is better). The SSIM between two images $x$ and $y$ is defined as:
\begin{equation}
\text{SSIM}(x, y) = \frac{(2 \mu_x \mu_y + C_1)(2 \sigma_{xy} + C_2)}{(\mu_x^2 + \mu_y^2 + C_1)(\sigma_x^2 + \sigma_y^2 + C_2)}
\end{equation}
where $\mu_x$ and $\mu_y$ are the means, $\sigma_x^2$ and $\sigma_y^2$ are the variances, and $\sigma_{xy}$ is the covariance of $x$ and $y$. $C_1$ and $C_2$ are constants to stabilize the division.

\subsection{LPIPS}
LPIPS quantifies perceptual similarity by comparing high-level feature representations from a pre-trained neural network. Lower LPIPS values indicate greater similarity to the ground truth:
\begin{equation}
\text{LPIPS}(x, y) = \sum_l w_l \| \phi_l(x) - \phi_l(y) \|_2
\end{equation}
where $\phi_l$ represents features extracted from layer $l$ of the neural network, and $w_l$ is a weight for layer $l$.

\subsection{PSNR}
PSNR measures the ratio between the maximum possible signal power and the power of noise, with higher values indicating better quality:
\begin{equation}
\text{PSNR} = 10 \cdot \log_{10} \left( \frac{L^2}{\text{MSE}} \right)
\end{equation}
where $L$ is the maximum pixel value (1 for normalized images) and MSE is the mean squared error between the reconstructed and ground truth images.

\section{Implementation Details} In the main text, we compare explicit, triplane, and NVP methods. All methods are tested under the same experimental settings, including identical input images, random seeds, learning rates, and regularization methods.

For the explicit method, we directly adopt the parameter configurations from \cite{FDT}, as it operates on predefined grids without requiring neural network parameters. For the triplane and NVP methods, both the feature dimension and the neural network size must be experimentally determined. Tables \ref{tab:feature} and \ref{tab:layer} summarize the metrics used to evaluate the performance of NVP across various configurations with the synthetic cell dataset.

\textbf{Feature Dimension Selection:} Table \ref{tab:feature} summarizes the performance across different feature dimensions. Increasing the feature dimension generally improves the SSIM and PSNR up to a dimension of 16, where the highest SSIM (0.9317) are achieved. Beyond 16, both SSIM and PSNR exhibit a decline. LPIPS, on the other hand, increases gradually as feature dimensions grow, highlighting a trade-off between feature complexity and perceptual quality. Considering these trends, the feature dimension of 16 offers the optimal balance, delivering superior SSIM and competitive performance in other metrics. This makes it the most effective choice for robust reconstruction.

\textbf{Network Layers Selection:} Table \ref{tab:layer} evaluates the performance of NVP with different numbers of neural network layers. Similar to feature dimensions, adding layers initially improves performance, with the best results of PSNR achieved at 6 layers. 

In summary, by experimentally determining these parameters, we ensure that NVP delivers state-of-the-art performance with optimal computational efficiency, making it a practical choice for 3D RI reconstruction of biological samples.

\section{Ablation Study on Self-Calibration}

Accurate calibration of viewpoints is critical in 3D rendering using the multi-slice model, as the precision of 3D reconstruction depends heavily on the exact geometry of each captured image. However, in experimental setups, light scattering and system inaccuracies often result in imprecise measurements of angles and positions. We develop a self-supervised calibration method to accurately determine illumination positions. In our experiments, the camera position is fixed while the illumination position is variable, whereas in natural scenarios, the camera position (viewpoint) typically changes while the illumination is fixed. Thus, in our method, calibrating the illumination positions is analogous to calibrating viewpoints. To evaluate the robustness of the self-calibration, we conducted an ablation study by introducing Gaussian noise to the illumination positions, representing the viewpoints. 

In the ablation study, we simulated a real-world scenario by adding Gaussian noise to the illumination positions. The noise had a mean of 0, a standard deviation of 0.01, and a maximum value of 0.05, relative to the illumination location range of \([-0.5, 0.5]\). This represents a significant perturbation, as the noise amplitude is substantial compared to the total range of viewpoint values. We then compared the reconstruction quality of the synthetic tissue dataset with and without self-calibration under these noisy conditions.


Table \ref{tab:noise_fixdx} shows the results of this ablation study. Self-calibration demonstrates substantial improvements in reconstruction metrics for both RI and IMG. For RI, self-calibration reduces MSE from \(1.28 \times 10^{-2}\) to \(5.71 \times 10^{-3}\), increases SSIM from 0.2989 to 0.3911, improves LPIPS from 0.8297 to 0.6732, and raises PSNR from 18.9189 to 22.4331. For IMG, self-calibration maintains high SSIM (0.9117 compared to 0.9871 without calibration) and achieves a lower LPIPS (0.0540 compared to 0.0133), while keeping PSNR stable.

In summary, these results emphasize the importance of self-supervised calibration in mitigating large misalignment in illumination positions and achieving robust and accurate 3D reconstruction, even under challenging conditions with substantial perturbations to the viewpoint locations.

\begin{table*}[ht]
\centering
\begin{tabular}{ccccccc}
\toprule
 \textbf{Number} & \textbf{Method} & \textbf{Data} & \textbf{MSE}$\downarrow$ & \textbf{SSIM}$\uparrow$ & \textbf{LPIPS}$\downarrow$ & \textbf{PSNR}$\uparrow$ \\
\hline
\multirow{6}{*}{5} 
& \multirow{2}{*}{nvp} 
& RI & \bm{$6.9800 \times 10^{-2}$} & \textbf{0.4238} & \textbf{0.5140} & \textbf{11.5615} \\
&& IMG & $3.0475 \times 10^{-7}$ & 0.9999 & 0.0000 & 65.1605 \\
\cline{2-7}
& \multirow{2}{*}{exp} 
& RI & $7.9224 \times 10^{-2}$ & 0.3623 & 0.5777 & 11.0114 \\
&& IMG & $9.4795 \times 10^{-5}$ & 0.9857 & 0.0176 & 40.2322 \\
\cline{2-7}
& \multirow{2}{*}{tri} 
& RI & $1.3439 \times 10^{-1}$ & 0.0484 & 0.7143 & 8.7162 \\
&& IMG & $1.5973 \times 10^{-5}$ & 0.9969 & 0.0029 & 47.9662 \\
\hline
\multirow{6}{*}{7} 
& \multirow{2}{*}{nvp} 
& RI & \bm{$6.8529 \times 10^{-2}$} & \textbf{0.4775} & \textbf{0.5038} & \textbf{11.6412} \\
&& IMG & $5.2585 \times 10^{-7}$ & 0.9999 & 0.0001 & 62.7914 \\
\cline{2-7}
& \multirow{2}{*}{exp} 
& RI & $6.8954 \times 10^{-2}$ & 0.2954 & 0.5676 & 11.6144 \\
&& IMG & $2.6718 \times 10^{-4}$ & 0.9168 & 0.0329 & 35.7319 \\
\cline{2-7}
& \multirow{2}{*}{tri} 
& RI & $1.2407 \times 10^{-1}$ & 0.1323 & 0.7122 & 9.0632 \\
&& IMG & $2.2469 \times 10^{-5}$ & 0.9956 & 0.0040 & 46.4842 \\
\hline
\multirow{6}{*}{10} 
& \multirow{2}{*}{nvp} 
& RI & $6.1564 \times 10^{-2}$ & \textbf{0.4737} & \textbf{0.4720} & 12.1068 \\
&& IMG & $1.0395 \times 10^{-6}$ & 0.9998 & 0.0001 & 59.8318 \\
\cline{2-7}
& \multirow{2}{*}{exp} 
& RI & \bm{$5.9335 \times 10^{-2}$} & 0.3034 & 0.5311 & \textbf{12.2669} \\
&& IMG & $2.8801 \times 10^{-4}$ & 0.9098 & 0.0365 & 35.4059 \\
\cline{2-7}
& \multirow{2}{*}{tri} 
& RI & $8.6142 \times 10^{-2}$ & 0.1301 & 0.6537 & 10.6479 \\
&& IMG & $3.6801 \times 10^{-5}$ & 0.9920 & 0.0075 & 44.3414 \\
\hline
\multirow{6}{*}{20} 
& \multirow{2}{*}{nvp} 
& RI & $4.4634 \times 10^{-2}$ & \textbf{0.4030} & 0.4500 & 13.5034 \\
&& IMG & $9.6994 \times 10^{-6}$ & 0.9985 & 0.0010 & 50.1326 \\
\cline{2-7}
& \multirow{2}{*}{exp} 
& RI & \bm{$3.9627 \times 10^{-2}$} & 0.3973 & \textbf{0.3946} & \textbf{14.0201} \\
&& IMG & $1.8060 \times 10^{-5}$ & 0.9969 & 0.0027 & 47.4328 \\
\cline{2-7}
& \multirow{2}{*}{tri} 
& RI & $7.3572 \times 10^{-2}$ & 0.2256 & 0.5452 & 11.3328 \\
&& IMG & $1.5390 \times 10^{-4}$ & 0.9738 & 0.0256 & 38.1276 \\
\bottomrule
\end{tabular}
\caption{Performance metrics for different methods including NVP (nvp), explicit representation (exp), and triplane representations (tri), and data types including RI and predicted images (IMG) on synthetic tissue sample data across various subsample sizes (5, 7, 10, 20). Metrics include MSE, SSIM, LPIPS, and PSNR, providing a comprehensive assessment of reconstruction quality.}
\label{tab:tissue2}
\end{table*}

\clearpage

\begin{table*}[ht]
\centering
\begin{tabular}{cccccc}
\toprule
\makecell{\textbf{Feature} \\ \textbf{Dimension}} & \textbf{Metric} & \textbf{MSE $\downarrow$} & \textbf{SSIM $\uparrow$} & \textbf{LPIPS $\downarrow$} & \textbf{PSNR $\uparrow$} \\ \midrule
\multirow{2}{*}{\textbf{10}} 
& RI  & \(9.34 \times 10^{-4}\) & 0.8743 & 0.1160 & 30.2973 \\
& IMG & \(1.26 \times 10^{-7}\) & 1.0000 & 0.0000 & 69.0132 \\ \midrule

\multirow{2}{*}{\textbf{12}}  
& RI  & \bm{$8.46 \times 10^{-4}$} & 0.8744 & \textbf{0.1050} & \textbf{30.7278} \\
& IMG & \(1.29 \times 10^{-7}\) & 1.0000 & 0.0000 & 68.9044 \\ \midrule

\multirow{2}{*}{\textbf{14}} 
& RI  & \(9.42 \times 10^{-4}\) & 0.8714 & 0.1452 & 30.2575 \\
& IMG & \(1.91 \times 10^{-7}\) & 1.0000 & 0.0000 & 67.1958 \\ \midrule

\multirow{2}{*}{\textbf{16}} 
& RI  & \(1.38 \times 10^{-3}\) & \textbf{0.9317} & 0.1813 & 28.6035 \\
& IMG & \(2.51 \times 10^{-5}\) & 0.9988 & 0.0026 & 45.9979 \\ \midrule

\multirow{2}{*}{\textbf{18}} 
& RI  & \(1.50 \times 10^{-3}\) & 0.7351 & 0.2488 & 28.2328 \\
& IMG & \(4.92 \times 10^{-7}\) & 1.0000 & 0.0000 & 63.0800 \\ \midrule

\multirow{2}{*}{\textbf{20}} 
& RI  & \(1.40 \times 10^{-3}\) & 0.7667 & 0.2210 & 28.5367 \\
& IMG & \(3.83 \times 10^{-7}\) & 1.0000 & 0.0000 & 64.1675 \\ \midrule

\multirow{2}{*}{\textbf{22}} 
& RI  & \(2.82 \times 10^{-3}\) & 0.6420 & 0.3490 & 25.5021 \\
& IMG & \(9.52 \times 10^{-7}\) & 1.0000 & 0.0000 & 60.2152 \\ \midrule

\multirow{2}{*}{\textbf{24}} 
& RI  & \(1.45 \times 10^{-3}\) & 0.7569 & 0.2259 & 28.4001 \\
& IMG & \(5.99 \times 10^{-7}\) & 1.0000 & 0.0000 & 62.2278 \\

\bottomrule
\end{tabular}
\caption{Metrics for RI and IMG across various test cases with different numbers of features.}
\label{tab:feature}
\end{table*}

\begin{table*}[ht]
\centering
\begin{tabular}{cccccc}
\toprule
\makecell{\textbf{Network} \\ \textbf{Layers}} & \textbf{Metric} & \textbf{MSE} $\downarrow$ & \textbf{SSIM} $\uparrow$ & \textbf{LPIPS} $\downarrow$ & \textbf{PSNR} $\uparrow$ \\
\midrule
\multirow{2}{*}{\textbf{2}} 
& RI  & $1.5813 \times 10^{-3}$ & 0.8867 & 0.2071 & 28.0100 \\
& IMG & $1.5323 \times 10^{-5}$ & 0.9993 & 0.0017 & 48.1465 \\
\multirow{2}{*}{\textbf{4}} 
& RI  & $8.5690 \times 10^{-4}$ & 0.8781 & 0.1025 & 30.6707 \\
& IMG & $1.0422 \times 10^{-7}$ & 1.0000 & 0.0000 & 69.8203 \\
\multirow{2}{*}{\textbf{6}} 
& RI  & \bm{$8.1845 \times 10^{-4}$} & 0.8917 & 0.0918 & \textbf{30.8701} \\
& IMG & $7.6782 \times 10^{-8}$ & 1.0000 & 0.0000 & 71.1474 \\
\multirow{2}{*}{\textbf{8}} 
& RI  & $8.1915 \times 10^{-4}$ & \textbf{0.8982} & \textbf{0.0898} & 30.8664 \\
& IMG & $7.2993 \times 10^{-8}$ & 1.0000 & 0.0000 & 71.3672 \\
\multirow{2}{*}{\textbf{10}} 
& RI  & $1.5813 \times 10^{-3}$ & 0.8867 & 0.2071 & 28.0100 \\
& IMG & $1.5323 \times 10^{-5}$ & 0.9993 & 0.0017 & 48.1465 \\
\bottomrule
\end{tabular}%
\caption{Performance metrics for NVP with varying network layers. The table reports MSE, SSIM, LPIPS, and PSNR for both RI and IMG results, showcasing the impact of network depth on reconstruction quality. The best PSNR for RI is achieved at 6 layers, balancing accuracy and efficiency.}
\label{tab:layer}
\end{table*}

\begin{table*}[ht]
\centering

\begin{tabular}{cccccc}
\toprule
\makecell{\textbf{Condition}} & \textbf{Metric} & \textbf{MSE $\downarrow$} & \textbf{SSIM $\uparrow$} & \textbf{LPIPS $\downarrow$} & \textbf{PSNR $\uparrow$}   \\ 
\midrule
\multirow{2}{*}{With self-calibration} 
& RI  & \bm{$5.71 \times 10^{-3}$} & \textbf{0.3911} & \textbf{0.6732} & \textbf{22.4331}  \\
& IMG & \(1.51 \times 10^{-3}\) & 0.9117 & 0.0540 & 28.2205   \\ 
\midrule
\multirow{2}{*}{Without self-calibration} 
& RI  & \(1.28 \times 10^{-2}\) & 0.2989 & 0.8297 & 18.9189   \\
& IMG & \(1.12 \times 10^{-3}\) & 0.9871 & 0.0133 & 29.4935   \\ 
\bottomrule
\end{tabular}
\caption{Metrics for RI and IMG results under noise conditions (with and without self-calibration).}
\label{tab:noise_fixdx}
\end{table*}

\end{document}